\documentclass[12pt]{iopart}

\usepackage{amsfonts}
\usepackage{lscape}

\begin{document}

\begin{center}\Large   
The "topological" charge for the finite XX quantum chain
\\[1cm]
\normalsize
Ulrich Bilstein\footnote{E-mail address: \tt bilstein@theoa1.physik.uni-bonn.de}
and Vladimir Rittenberg\footnote{E-mail address: \tt vladimir@th.physik.uni-bonn.de}
\\[0.5cm]
 Universit\"{a}t Bonn \\
                    Physikalisches Institut, Nu\ss allee 12,
                    D-53115 Bonn, Germany\\[1.3cm]
\end{center} 
{\bf \small Abstract.}
\small
It is shown that an operator (in general non-local) commutes with the
Hamiltonian describing the finite XX quantum chain with certain non-diagonal
boundary terms. In the infinite volume limit this operator gives the
"topological" charge.
\\[0.2cm]
\normalsize
\quad\\[0.7cm]
We consider the $L$ sites XX quantum chain with non-diagonal boundary terms
given by the Hamiltonian:
\begin{equation}
H=\frac{1}{4}\sum_{j=1}^{L-1}
\left(\sigma_j^x\sigma_{j+1}^x+\sigma_j^y\sigma_{j+1}^y\right)
+ \lambda \sigma_1^x
+\mu \left[ \cos(\chi) \sigma_L^x - \sin(\chi)\sigma_L^y \right]
\label{HXX}
\end{equation}
here $\lambda,\mu$ and $\chi$ are parameters. Obviously the $O(2)$ symmetry on the bulk
terms is broken by the boundary terms.
The question is if this finite quantum chain has no "hidden" symmetries
(hidden symmetries occur normally in the thermodynamic limit only).
 We will show that it does. In order to
understand the problem and find the "hidden" symmetries, it is convenient \cite{JT} to add two more sites
denoted with $0$ and $L+1$ and consider another Hamiltonian that we denote by
\begin{equation}
\fl
H_{\rm f}=\frac{1}{4}\sum_{j=1}^{L-1}
\left(\sigma_j^x\sigma_{j+1}^x +\sigma_j^y\sigma_{j+1}^y\right)
+\lambda\sigma_0^x\sigma_1^x
+\mu\left[\cos(\chi)\sigma_L^x\sigma_{L+1}^x-\sin(\chi)\sigma_L^y\sigma_{L+1}^x\right] .
\label{Hf}
\end{equation}
(in Refs. \cite{BW,UB} $H_{\rm f}$ was denoted by $H_{\rm long}$).
 Notice that $H_{\rm f}$ is $Z(2)\times Z(2)$ symmetric since the matrices $\sigma^x_0$ and
$\sigma^x_{L+1}$ commute with $H_{\rm f}$. Therefore the spectrum of $H_{\rm f}$
decomposes into four blocks $(\pm,\pm),(\pm,\mp)$ corresponding to the eigenvalues $\pm 1$ of
the $\sigma^x$ matrices.
The Hamiltonian $H$ given by eq.\eref{HXX} corresponds to the $(+,+)$ block.
Actually the symmetry of $H_{\rm f}$ is larger than $Z(2)\times Z(2)$, it is the finite
non-abelian group with the three generators $\sigma^x_0, \sigma^x_{L+1}$ and
$U=\sigma_0^z\sigma_1^z\cdots\sigma_{L+1}^z$ (which also commutes with $H_{\rm f}$).
Moreover using the obvious relations:
\begin{equation}
\{ \sigma_0^x, U \}=\{ \sigma_{L+1}^x, U \}=0; \qquad U^2=1
\end{equation}
one can show that if $|\mu,\nu,E\rangle $ is an eigenvector of $H_{\rm f}$ in the $(\mu,\nu)$
sector, corresponding to the eigenvalue $E$, then
\begin{equation}
U\mid \mu,\nu,E\rangle =\mid -\mu,-\nu,E\rangle 
\end{equation}
is also an eigenvector of $H_{\rm f}$, in the $(-\mu,-\nu)$ sector corresponding to the
same eigenvalue $E$. As a result,
 the whole
spectrum of $H_{\rm f}$ is doubly degenerate: the spectra in the sectors $(+,+)$ and $(-,-)$
(respectively $(+,-)$ and $(-,+)$) are identical. Since the degeneracy includes
the ground-state, the symmetry is spontaneously broken.
 The Hamiltonian $H_{\rm f}$ can be diagonalized in terms of free fermions which does
not mean that the fermionic energies or the wave functions are simple to
derive since the secular equation is hard to solve. This problem is discussed
in detail in Ref.\cite{BW}. Moreover, for our purpose one needs to find boundary
terms which lead to fermionic energies of a special form (the reason is
going to be explained shortly), therefore from our collection of
 boundary terms where we could find solutions of the
secular equation, we will consider only two kinds 
of boundary conditions:
\begin{itemize}
\item{
\begin{equation}
\label{type1}
\fl
\lambda=\mu=\frac{1}{\sqrt{8}}
\end{equation}
}
\item{
\begin{equation}
\label{type2}
\fl
\lambda=\mu=\frac{1}{4} ; \qquad \chi=0
\end{equation}
}
\end{itemize}
We have to consider the cases $L$ odd and even separately. The reason is the
following one: for $L$ odd the ground-state of $H_{\rm f}$ is in the $(+,+)$ sector in
which we are interested. For $L$ even the ground-state is in the $(+,-)$ sector.
Since we are more interested in the spectra where the ground-state is, we
will confine ourselves to the case $L$ odd
(as opposed to the case of diagonal
boundary terms where the ground-state appears in chains with an even number of sites \cite{AB}).
One can repeat the whole procedure described below for $L$ even.
 For both cases \eref{type1} and \eref{type2} one finds a zero fermionic mode which explains 
in a different way
why each level of $H_{\rm f}$ is doubly degenerate. We now specialize to the
boundary condition \eref{type1}. The energy gaps are given \cite{BW} by the fermionic
energies (the zero mode not included)
\begin{equation}
\label{fen}
\Lambda_n^{\pm}=
\sin\left(\frac{2n+1}{L+1}\frac{\pi}{2} \pm \frac{\chi}{L+1}\right) \qquad 0\leq n \leq (L-1)/2
\end{equation}
It is convenient to consider the operator
\begin{equation}
\label{tdef}
T_{\rm f}=\frac{1}{2}\sum_{n=0}^{(L-1)/2} (N^+_n-N^-_n) .
\end{equation}
where $N_n^{\pm}$ are fermionic number operators ($N_n^{\pm}=0,1$) corresponding
to the energies given by eq.\eref{fen}. The eigenvalues of $T_{\rm f}$ denoted by $m$
are obviously given by integer or half-integer numbers. Obviously this
allows a $Z(2)$ grading. It turns out \cite{BW} that in the $(+,+)$ sector $m$ takes
integer values (even number of fermions) whereas in the $(+,-)$ sector it
takes half-integer values (odd number of fermions). 
 We define the partition function corresponding to the eigenvalue $m$ of
$T_{\rm f}$, for a chain with $L$ sites
\begin{equation}
Z_m(L)=\tr z^{\frac{L}{\pi}\left[ \sum_{n=0}^{(L-1)/2} 
\left( \Lambda_n^+ N_n^+ +\Lambda_n^- N_n^-\right) \right]  }
\end{equation}
with the constraint 
\begin{equation}
\frac{1}{2}\sum_{n=0}^{(L-1)/2} (N^+_n-N^-_n)=m
\end{equation}
and consider the sectors $(+,+)$ and $(+,-)$ only.
In the thermodynamic limit one obtains:
\begin{equation}
\label{partm}
Z_m=\lim_{L\to \infty} Z_m(L)
= z^{2(m+\chi/(2\pi))^2-\chi^2/(2\pi^2)}
\prod_{n=1}^{\infty}(1-z^n)^{-1}
\end{equation}
which is what one would expect if $T_{\rm f}$ corresponds to the "topological" charge.
If we consider the $(+,+)$ sector (which as mentioned corresponds to the
original Hamiltonian given by eq.\eref{HXX}) and sum over $m$ integer the partition
functions given by eq.\eref{partm}, one gets the known partition function for a
compactified bosonic field (with the correct radius) and von Neumann
boundary conditions at both ends \cite{UB,OA,HS}. Actually the whole algebraic
structure of the problem, in the thermodynamic limit, can be read-off from the
expressions \eref{fen}: 
\begin{equation}
\lim_{L\to \infty} \left( \frac{L}{\pi}\Lambda_n^{\pm}\right)=n+1/2\pm\chi/\pi
\end{equation}
if we keep in mind that the sum of two integer numbers is an integer number.
One could pursue this issue further and make contact with the sine-Gordon
model with a boundary at the free fermion point where much work was done
\cite{GZ,AML,MN}. In this paper we are however interested in the properties
of the finite chain given by eq.\eref{HXX} and want therefore to get the
"topological" charge $T$ for this chain. In order to do so, we have to write
$T_{\rm f}$ given by eq.\eref{tdef} in the basis where $H_{\rm f}$ is written (see eq.\eref{Hf}) and
project on the $(+,+)$ sector. This is a lengthy calculation where we have
used the results of Ref.\cite{BW}. One obtains
\begin{eqnarray}
\fl
T=\frac{-1}{4L+4} \sum_{\stackrel{k,j=1}{\stackrel{k+j {\rm\, odd}}{\scriptscriptstyle k<j}}}^L
\Bigg\{\nonumber \\
\left[ f(j-k)\cos(\chi\frac{k-j}{L+1})-(-1)^k f(k+j) \cos(\chi \frac{k+j}{L+1}) \right] 
\sigma_k^y \sigma_{k+1}^z\cdots \sigma_{j-1}^z \sigma_j^x
\nonumber\\
-\left[ f(j-k)\cos(\chi\frac{k-j}{L+1})+(-1)^k f(k+j) \cos(\chi \frac{k+j}{L+1}) \right]
\sigma_k^x \sigma_{k+1}^z\cdots \sigma_{j-1}^z \sigma_j^y
\nonumber\\
+\left[ f(j-k)\sin(\chi\frac{k-j}{L+1})+(-1)^k f(k+j) \sin(\chi \frac{k+j}{L+1}) \right] 
\sigma_k^y \sigma_{k+1}^z\cdots \sigma_{j-1}^z \sigma_j^y
\nonumber\\
+\left[ f(j-k)\sin(\chi\frac{k-j}{L+1})-(-1)^k f(k+j) \sin(\chi \frac{k+j}{L+1}) \right]
\sigma_k^x \sigma_{k+1}^z\cdots \sigma_{j-1}^z \sigma_j^x
\Bigg\}
\nonumber \\
+\frac{1}{\sqrt{8}(L+1)}  \sum_{\stackrel{j=1}{\scriptscriptstyle j {\rm\, odd}}}^L
\Bigg\{ f(j) \cos(\chi \frac{j}{L+1})  \sigma_{1}^z\cdots \sigma_{j-1}^z \sigma_j^y
\nonumber \\
f(j)\sin(\chi \frac{j}{L+1})  \sigma_{1}^z\cdots \sigma_{j-1}^z \sigma_j^x
-f(j+L+1)\cos(\chi \frac{j}{L+1})  \sigma_j^y \sigma_{j+1}^z\cdots \sigma_{L}^z
 \nonumber \\
-f(j+L+1)\sin(\chi \frac{j}{L+1})  \sigma_j^x \sigma_{j+1}^z\cdots \sigma_{L}^z
\Bigg\}
\label{horr}
\end{eqnarray}
where
\begin{equation}
f(x)=1/\sin\left( \frac{x\pi}{2L+2} \right). 
\end{equation}
$T$ is a pseudoscalar for $\chi=0$ when $H$ is parity invariant. One can also check that
\begin{equation}
[T,H]=0
\end{equation} 
only for $L$ odd and not for $L$ even. The expression \eref{horr} is horrible, what
is important is that it exists. 
Let us stress that we were able to identify $T_{\rm f}$ and therefore $T$ only
because the fermionic energies had the form \eref{fen}. We have not found \cite{BW}  any
other boundary conditions where we know the spectrum of the finite
chain and where the partition functions have the form given by eq.\eref{partm}
with $\chi$ different of zero (only in this case we can write $T_{\rm f}$ like in
eq.\eref{tdef}). 
We would have liked to have more examples in order to make sure that the 
boundary condition \eref{type1} is not a special case and that only for this case 
one can find the "topological" charge.
There is however one case (given by the boundary condition \eref{type2})
where although the spectrum of the finite quantum chain $H$ is twice degenerate 
(nothing to do with a zero mode but with the fact that in the continuum it
gives the partition function \eref{partm} with $\chi=0$):
\begin{equation}
\label{el}
\Lambda_n^+=\Lambda_n^-=\sin \left(\frac{2n+1}{L+2}\frac{\pi}{2}\right)\qquad 0\leq n \leq (L-1)/2
\end{equation}
we were able to identify the "topological" charge. We did it using a trick.
We took $\lambda=\mu=1/4$ and a small value for $\chi$ in eq.\eref{Hf}, diagonalized
 $H_{\rm f}$ numerically and identified $\Lambda^{\pm}_n$ and the creation and
annihilation operators corresponding to these energy levels.
 This has allowed us to guess $T_{\rm f}$
for $\chi=0$ where the spectrum of $H_{\rm f}$ is known. We found:
\begin{equation}
\label{Tf2}
\fl
T_{\rm f}=
 \frac{1}{8}\sum_{j=1}^{L-1} \left[
\left(1+(-1)^j\right) \sigma_j^x \sigma_{j+1}^y -\left(1-(-1)^j\right)\sigma_j^y\sigma_{j+1}^x
\right]
+\frac{1}{4}\left(\sigma_0^x\sigma_1^y -\sigma_L^y\sigma_{L+1}^x\right)
\end{equation}
and 
\begin{eqnarray}
\label{T2}
\fl
T = \frac{1}{8}\sum_{j=1}^{L-1} \left[
\left(1+(-1)^j\right) \sigma_j^x \sigma_{j+1}^y -\left(1-(-1)^j\right)\sigma_j^y\sigma_{j+1}^x
\right]
+\frac{1}{4}\left(\sigma_1^y -\sigma_L^y\right) .
\end{eqnarray}
Notice that the expressions \eref{Tf2} and \eref{T2} are local ones. One can look at
$T$ and consider it as a quantum chain (keep in mind that it commutes with $H$
only for $L$ odd). This quantum chain has amusing properties. If we
disregard the boundary terms, it is trivial to diagonalize it and one obtains
\begin{equation}
\frac{1}{2}\left(\sum_{k=1}^{L-1}N_k\right) -\frac{L-1}{4}
\end{equation}
here $N_k$ are fermionic operators. If one takes into account the boundary
terms, in order to diagonalize it, one has to use $T_{\rm f}$ given by eq.\eref{Tf2} and
project in order to find the spectrum of $T$ (the way we did in order to find
the spectrum of $H$ out of the spectrum of $H_{\rm f}$). For $L$ odd the spectrum is
the known one (integer values) since $T$ is the "topological" charge.
 For $L$ even, the spectrum is complicated.

 One can ask what we have learned from our exercise. The fact that the
quantum chain \eref{HXX} has many conservation laws (the total number of
fermions is just one example) should not be a surprise
since its spectrum is related to the one of $H_{\rm f}$ which is a free system. We
think that the fact that we have been able to identify the "topological"
charge which is related to the magnetic charge in the Coulomb gas
description or to vortices (see \cite{BN} for a review on the subject) on a
finite lattice is interesting and that this identification can  probably be done not only for the XX
chain, but for the XXZ chain too. 
How to do this generalization is by no means clear to us.


\ack{ We would like to thank F.C.Alcaraz, M.Fabrizio, G.Mussardo, A.Nersesyan and
P.Simon for useful discussions. V.R. would like to thank SISSA for
hospitality under the TMR Grant ERBFMRXCT 960012.}

\end{document}